\documentclass[a4j,11pt]{article}

\usepackage[dvipdfmx]{graphicx}
\usepackage{here}
\usepackage{dcolumn}
\usepackage{bm}
\usepackage{amsmath}
\usepackage{amssymb}
\usepackage{mathrsfs}
\usepackage{color}
\usepackage[subrefformat=parens]{subcaption}
\usepackage{authblk}

\newtheorem{remark}{{\bf Remark}}

\def \rmd{\mathrm{d}}

\newcommand{\qed}{\nobreak \ifvmode \relax \else
     \ifdim\lastskip<1.5em \hskip-\lastskip
     \hskip1.5em plus0em minus0.5em \fi \nobreak
     \vrule height0.75em width0.5em depth0.25em\fi}

\begin{document}
\title{Kinetic construction of the high-beta anisotropic-pressure equilibrium in the magnetosphere}
\author[1]{H. Aibara}
\author[1,2]{Z. Yoshida}
\author[1]{K. Shirahata}
\affil[1]{Graduate School of Frontier Sciences, The University of Tokyo, Kashiwa, Chiba 277-8561, Japan}
\affil[2]{National Institute for Fusion Science, Toki, Gifu 509-5292, Japan}
\date{\today}

\maketitle

\begin{abstract}
A theoretical model of the high-beta equilibrium of magnetospheric plasma was constructed by consistently connecting the (anisotropic pressure) Grad--Shafranov equation and the Vlasov equation.
The Grad--Shafranov equation was used to determine the axisymmetric magnetic field for a given magnetization current corresponding to a pressure tensor. 
Given a magnetic field, we determine the distribution function as a specific equilibrium solution of the Vlasov equation, using which we obtain the pressure tensor. 
We need to find an appropriate class of distribution function for these two equations to be satisfied simultaneously.
Here, we consider the distribution function that maximizes the entropy on the submanifold specified by the magnetic moment.
This is equivalent to the reduction of the canonical Poisson bracket to the noncanonical one having the Casimir corresponding to the magnetic moment.
The pressure tensor then becomes a function of the magnetic field (through the cyclotron frequency) and flux function, satisfying the requirement of the Grad--Shafranov equation. 
\end{abstract}


\section{Introduction}
The magnetosphere is a naturally made system confining a high-beta plasma\,\cite{krimigis1979}. 
A similar system may be created for fusion energy applications\,\cite{hasegawa1987,hasegawa1990}. 
Laboratory magnetospheres, namely, LDX\,\cite{garnier2009}
and RT-1\,\cite{yoshida2006},
demonstrated a stable confinement of high-beta ($\sim$ 1) plasmas.

The aim of this study is to formulate a theoretical model of high-beta equilibrium in the magnetosphere.
Magnetized particles in an axisymmetric magnetic field have three independent first integrals, facilitating an easy construction of the equilibrium solutions of the drift kinetic equation.
The strong inhomogeneity of the dipole magnetic field is the key to understanding the localization of magnetized particles to the vicinity of the magnetic dipole\,\cite{hasegawa1987}.
However, in a high-beta plasma, the magnetic field must be corrected by considering the diamagnetic current.
We observe a significant expansion of the dipole field due to plasma pressure, which can be used to estimate the plasma pressure\,\cite{saitoh2011,yoshida2012}.
At the first-order level, we may invoke the Grad--Shafranov equation\,\cite{grad1958} to analyze the magnetic field of a finite-beta plasma.
However, this equation falls short of considering the strong anisotropy of the distribution function, as the kinetic model makes predictions for the magnetospheric system.
The bouncing particles introduces variations in the velocity distribution function along the field lines.
The anisotropic temperature is also demonstrated experimentally in RT-1\,\cite{kawazura2015}. 
Appropriate corrections can be made by using the extended Grad--Shafranov equation, which was developed to model the equilibrium of mirror systems\,\cite{grad1967proceedings,grad1967}.
In fact, each magnetic flux tube in the magnetosphere may be viewed as a crescent-shaped mirror system.
The extended Grad--Shafranov equation, which is still a macroscopic magneto-fluid model, considers an anisotropic pressure that is a function of the two-dimensional magnetic coordinates, namely, the flux function and magnetic field strength.
While the functional form of the pressure tensor remains arbitrary in such a fluid model, parametric studies on the effect of anisotropic pressure have been performed using numerical analysis\,\cite{krasheninnikov2000,Furukawa}. 
We have yet to build a consistent relationship between the kinetic description and the magneto-fluid model, and to provide a physical reason for selecting an appropriate form of the pressure tensor.

In the present study, we constructed a self-consistent model based on the idea of the maximum entropy state in a topologically constrained phase space (or a symplectic leaf foliated by Casimirs)\,\cite{yoshida2014self}.
 In the context of magnetospheric plasma confinement, the adiabatic invariant acts as a topological constraint (Casimir of the noncanonical Hamiltonian mechanics\,\cite{Morrison}). 
Providing the topological charge (in fact, the Casimir) with a chemical potential, we define a grand canonical ensemble, on which we consider the Gibbs distribution.
As the magnetic moment is the relevant Casimir, the pair of chemical potential and Casimir parallels that of the magnetic field and magnetization in the well-known model of magnetic materials.
Such a ``thermal equilibrium'' yields the desired pressure tensor to be used in the generalized Grad--Shafranov equation.

The remainder of this paper is organized as follows:
In the following section, we review the Hamiltonian mechanics of magnetized particles and derive the stationary distribution function that describes the thermal equilibrium under the topological constraint given by the magnetization.  
The corresponding pressure tensor is used to formulate the generalized (anisotropic pressure) Grad--Shafranov equation in  Sec.\,\ref{sec:Magnetic_field}.
In Sec.\,\ref{sec:result}, we show examples of numerical solutions, as well as some relations useful to estimate the anisotropic pressure effect.
Section\,\ref{sec:conclusion2} concludes the paper.

\section{Particle motion and magnetic field}
\label{subsec:Particle_motion}
To construct the high-beta equilibrium of magnetospheric plasma, we combine two models:
one is the kinetic model for calculating the distribution function,
and the other is the macroscopic magneto-fluid magnetohydrodynamics (MHD) model for calculating the magnetic field. 
The distribution function is given as a stationary solution of the Vlasov theory with appropriate coarse graining.
Evaluating the pressure tensor using the distribution function, we solve the generalized (anisotropic pressure) Grad--Shafranov equation to determine the magnetic field.

\subsection{Hamiltonian of magnetized particles in magnetosphere}
\label{subsec:Kinetic}
In an axisymmetric magnetic field (of sufficient strength), the dynamics of a magnetized particle consist of three different periodic motions: gyro motion, bounce motion, and drift motion. 
It is then convenient to span the phase space by variables  
\begin{equation}
    \bm{z}=(\theta_g, \, \mu, \, ; \, \ell , \,  P_\parallel \, ;\, \theta, \, P_\theta ),\label{coordinate}
\end{equation}
where $\mu:=J_gq/m$ is the magnetic moment ($J_g$ is the action corresponding to the gyro motion,
$q$ is the particle charge, and $m$ is the particle mass), $\theta_g$ is the gyro angle, $P_\parallel$ is the canonical momentum parallel to the magnetic field, $\ell$ is the parallel coordinate that constitutes the canonical pair with $P_\parallel$, $P_\theta$ is the canonical angular momentum around the geometrical axis, and $\theta$ is the azimuthal angle.
(For convenience, we use $\ell , \,  P_\parallel$, instead of the bounce action-angle pair).
Neglecting the kinetic part of the canonical angular momentum,
we approximate $P_\theta=q\psi$
($\psi=rA_\theta$ is the flux function in the cylindrical coordinate system $(r,\theta,z)$, where $A_\theta$ is the $\theta$ component of the vector potential). 

Here, we consider two different ``reductions'' for the Hamiltonian.
First, the gyro angle $\theta_g$ is \emph{coarse-grained} and is eliminated from the Hamiltonian (such a Hamiltonian only dictates the guiding center motion of the magnetized particle). 
Then, we obtain
\begin{equation}
    \dot{\mu}=\frac{\partial H}{\partial \theta_g}=0. \label{Hamiltonian-condition1}
\end{equation}
Second, the Hamiltonian is independent of the azimuthal angle $\theta$ because we consider an axisymmetric magnetic field.
With the approximation $P_\theta=q\psi$, we obtain 
\begin{equation}
    q\dot{\psi}=\dot{P_\theta}=\frac{\partial H}{\partial \theta}=0. \label{Hamiltonian-condition2}
\end{equation}
The two constants $\mu$ and $\psi$ play an important role in later discussion. 

The magnetic field may be written as $\bm{B}=\nabla\psi\times\nabla\theta$ (note that the toroidal magnetic field is absent in the magnetospheric system). 
By adding the coordinate $\ell$ that measures the arc length on each magnetic field line, we define a magnetic coordinate system $(\ell,\psi,\theta)$. 
Through the axisymmetry, we can eliminate $\theta$. 
The reader is referred to Refs.\,\cite{hasegawa2005motion,yoshida2016self} for the parameterization of magnetized particles in a dipole magnetic field.

\begin{remark}[The adiabatic invariant as a Casimir invariant]
\label{remark:Casimr}
\normalfont
In the view of the Poisson bracket, coarse-graining of the gyro angle is represented by the modified Poisson matrix\,\cite{yoshida2014self}
\begin{equation}
\mathcal{J}_\mu=
\begin{pmatrix}
    0 & 0 & 0 \\
    0 & J_c & 0 \\
    0 & 0 & J_c \\
\end{pmatrix}, \:
J_c=
\begin{pmatrix}
    0 & 1 \\
    -1 & 0 \\
\end{pmatrix}.
\end{equation}
Here, we consider the variables (\ref{coordinate}). 
Then, the modification of the Poisson matrix makes the Poisson bracket
\begin{equation}
    \{F,G\}_\mu:=\langle\partial_zF,\,\mathcal{J}_\mu\partial_zG\rangle
\end{equation}
noncanonical and the adiabatic invariant $\mu$ is a Casimir invariant. 
The foliation of the phase space induced by the Casimir invariant is essential in our theory.
\end{remark}

\subsection{Kinetic distribution function}
\label{subsec:distribution}
\subsubsection{General form of stationary distribution function}
In the Vlasov theory, the stationary distribution function $f$ is given by
\begin{equation}
\{ H, f^* \} =0,
\label{Vlasov}
\end{equation}
where $\{~,~\}$ is the canonical Poisson bracket, and $f^*$ is the Hodge dual of $f$, 
i.e., $f= f^* \mathrm{vol}$ with the phase-space volume element $\mathrm{vol}=\rmd^3x \rmd^3v$.
When $\{ G_j ,H\}=0$, $G_j$ is a constant of motion.
A scalar function such as $f^*(H,G_1,\cdots,G_n)$ gives a stationary distribution $f=f^*\mathrm{vol}$,
because
\[
\{ H, f^*(H,G_1,\cdots,G_n) \} = \frac{\partial f^*}{\partial H} \{ H, H \}
+ \sum_{j=1}^m \frac{\partial f^*}{\partial G_j} \{ G_j, H \} =0.
\]
Here, we use the aforementioned $\mu$ and $\psi$ to define a stationary distribution function using \begin{equation}
    f^*=g(H,\mu,\psi),
\end{equation}
where $g$ is an arbitrary function of $H$, $\mu$, and $\psi$.

\subsubsection{Thermal equilibrium with topological constraints}\label{Maxwellian}
Although the collisionless kinetic theory leaves infinite freedom in the stationary distribution function, it provides us with the theoretical basis for statistical mechanics to define the most probable distribution with the appropriately defined entropy, that is, the invariant measure is determined by the guide of the Poisson structure pertinent to the kinetic theory.
Here, the invariant measure is given on the symplectic leaf, defined as the level sets of the two invariants $\mu$ and $\psi$.

We consider the Hamiltonian 
\begin{equation}
    H=\mu B(\ell,\psi)+\frac{P_\parallel^2}{2m}, \label{Hamiltonian-1}
\end{equation}
where $B$ is the magnetic field. 
Here, we consider a quasi-neutral plasma ($\phi=0$) and neglect the kinetic energy of the toroidal drift velocity by approximating $P_\theta=q\psi$. 
The first term $\mu B$ may be regarded as the potential energy ($\mu$ is constant for each particle) on each contour of $\psi$ (i.e., magnetic field line). 

Leaving only $\psi$ as a free parameter characterizing the thermal nonequilibrium of the system, we consider the thermal equilibrium (maximum entropy) distribution function such that
(see Appendix\,\ref{appendix:GCE})
\begin{equation}
    f^*=A(\psi)\exp\biggl(-\frac{H}{T_{\parallel 0}}\biggr)\exp\biggl(-\mu B_0(\psi)\frac{T_{\parallel 0}-T_{\perp 0}}{T_{\parallel 0} T_{\perp 0}}\biggr),\label{f-star1}
\end{equation}
where the two constants $T_{\parallel 0}$ and $T_{\perp 0}$ represent the parallel and perpendicular temperatures at $\ell=0$, respectively.
We also define $B_0(\psi):=B(0,\psi)$, where $B(\ell,\,\psi)$ is the magnetic field strength evaluated as a function of $\ell$ and $\psi$. 
Then, by introducing $v_\parallel$ and $v_\perp$ (which represent the parallel and perpendicular velocity with respect to the magnetic field) and substituting (\ref{Hamiltonian-1}) and $\displaystyle\mu=\frac{mv_\perp^2}{ 2 B (\ell,\,\psi)}$ in (\ref{f-star1}), we can rewrite (\ref{f-star1}) as
\begin{equation}
    f^*=A(\psi)\exp\biggl\{-\biggl(\frac{mv_\parallel^2}{2T_\parallel(\ell,\psi)}+\frac{mv_\perp^2}{2T_\perp(\ell,\psi)}\biggr)\biggr\}, 
\end{equation}
where
\begin{eqnarray}
    T_\parallel(\ell,\psi)&&=T_{\parallel 0},\label{t-parallel}\\
    T_\perp(\ell,\psi)&&=\frac{T_{\parallel 0}T_{\perp 0}}{T_{\perp 0}+\frac{B_0(\psi)}{B(\ell,\psi)}(T_{\parallel 0}-T_{\perp 0})}.\label{t-perp}
\end{eqnarray}
Therefore, our distribution function is a Maxwellian with varying anisotropic temperatures (\ref{t-parallel}) and (\ref{t-perp}).
Multiplying the phase-space volume form, we obtain
\begin{eqnarray}
    f&&=f^*\rmd^3x\,\rmd^3v\\
    &&=f^*\rmd^3x\,\frac{D(v_x,v_y,v_z)}{D(v_\parallel,v_\perp,\theta_g)}\rmd v_\parallel \rmd v_\perp \rmd\theta_g\\
    &&=v_\perp A(\psi)\exp\biggl\{-\biggl(\frac{mv_\parallel^2}{2T_\parallel(\ell,\psi)}+\frac{mv_\perp^2}{2T_\perp(\ell,\psi)}\biggr)\biggr\}\rmd^3x\,\rmd v_\parallel \rmd v_\perp \rmd\theta_g.\nonumber\\
\end{eqnarray}
Transforming variables in the volume form, we rewrite
\begin{eqnarray}
    f&&(\ell,\psi,v_\parallel,v_\perp)=\nonumber \\
    &&2\pi v_\perp A(\psi)\exp\biggl\{-\biggl(\frac{mv_\parallel^2}{2T_\parallel(\ell,\psi)}+\frac{mv_\perp^2}{2T_\perp(\ell,\psi)}\biggr)\biggr\}\rmd^3x\,\rmd v_\parallel \rmd v_\perp.\nonumber\\
\end{eqnarray}

The corresponding configuration space density is 
\begin{eqnarray}
    n(\ell,\psi)&&=\iint f(\ell,\psi,v_\parallel,v_\perp)\, \rmd v_\parallel \rmd v_\perp \\
    &&=A(\psi)\biggl(\frac{2\pi}{m}\biggr)^{\frac{3}{2}}T_\parallel^{\frac{1}{2}}(\ell,\psi)\,T_\perp(\ell,\psi),
\end{eqnarray}
and the parallel and perpendicular components of the pressure tensor are 
\begin{eqnarray}
    p_\parallel(\ell,\psi)&&=\iint f(\ell,\psi,v_\parallel,v_\perp)\, m v_\parallel^2 \rmd v_\parallel \rmd v_\perp\\
    &&=A(\psi)\biggl(\frac{2\pi}{m}\biggr)^{\frac{3}{2}}T_\parallel^{\frac{3}{2}}(\ell,\psi)T_\perp(\ell,\psi),\\
    p_\perp(\ell,\psi)&&=\frac{1}{2} \iint f(\ell,\psi,v_\parallel,v_\perp)\, m v_\perp^2 \rmd v_\parallel \rmd v_\perp\\
    &&=A(\psi)\biggl(\frac{2\pi}{m}\biggr)^{\frac{3}{2}}T_\parallel^{\frac{1}{2}}(\ell,\psi)T_\perp^{2}(\ell,\psi).
\end{eqnarray}
Finally, we represent the pressures and functions of the magnetic coordinates $\psi$ and $B$, and using (\ref{t-parallel}) and (\ref{t-perp}), we obtain
\begin{eqnarray}
    p_\parallel(\psi,B)&&=A(\psi)\biggl(\frac{2\pi}{m}\biggr)^{\frac{3}{2}}T_{\parallel 0}^{\,\frac{5}{2}}\,\frac{T_{\perp 0}}{T_{\perp 0}+\frac{B_0(\psi)}{B}(T_{\parallel 0}-T_{\perp 0})},\label{p_parallel}\\
    p_\perp(\psi,B)&&=A(\psi)\biggl(\frac{2\pi}{m}\biggr)^{\frac{3}{2}}T_{\parallel 0}^{\,\frac{5}{2}}\biggl(\frac{T_{\perp 0}}{T_{\perp 0}+\frac{B_0(\psi)}{B}(T_{\parallel 0}-T_{\perp 0})}\biggr)^2.\label{p_perp}\nonumber\\
\end{eqnarray}

\section{Finite-beta equilibrium}
\label{sec:Magnetic_field}
The anisotropic pressures (\ref{p_parallel}) and (\ref{p_perp}), evaluated for the thermal equilibrium on the symplectic leaf of the magnetic moment, can now be used in the generalized Grad--Shafranov equation.
The pressure tensor is  
\begin{equation}
\bm{P} = \bm{b}\bm{b}p_\parallel + (\bm{I}-\bm{b}\bm{b})p_\perp\label{pressure-tensor}
\end{equation}
where $p_\parallel$ and $p_\perp$ are the parallel and perpendicular pressures, respectively, and $\bm{b}$ is the unit vector parallel to $\bm{B}$. 
The MHD equilibrium equations are
\begin{eqnarray}
(\nabla \times\bm{B}) \times \bm{B} = \mu_0 \nabla\cdot\bm{P},&&\\
\nabla \cdot \bm{B} = 0,&&
\end{eqnarray}
where $\mu_0$ is the vacuum permeability. 
In the axisymmetric magnetospheric system (which has no toroidal magnetic field), we can convert the MHD equilibrium equation into a generalized (anisotropic pressure) Grad--Shafranov equation\,\cite{grad1967proceedings,grad1967}:
\begin{equation}
\Delta^*\psi = -\mu_0 r^2\left.\frac{\partial p_\parallel}{\partial \psi}\right|_B - \frac{1}{\sigma}\nabla\psi\cdot\nabla\sigma\label{GS-1}
\end{equation}
\begin{equation}
\left.\frac{\partial p_\parallel}{\partial B}\right|_\psi + \frac{p_\perp-p_\parallel}{B} = 0,\label{GS-2}
\end{equation}
where $\Delta^*$ denotes the Grad--Shafranov operator
\begin{equation}
\Delta^*:=r^2\nabla\cdot(\frac{1}{r^2}\nabla), 
\end{equation}
and $\sigma$ is defined as 
\begin{equation}
\sigma:=1+\mu_0\frac{p_\perp -p_\parallel}{B^2}.
\end{equation}
We notice that the equilibrium equations (\ref{GS-1}) and (\ref{GS-2}) can also be obtained as equilibrium equations of the CGL-MHD model\cite{chew1956boltzmann}.

Note that (\ref{p_parallel}) and (\ref{p_perp}) satisfy the second equation (\ref{GS-2}).
We only need to solve (\ref{GS-1}) to determine $\psi$.

\section{Numerical analysis}
\label{sec:result}
In this section, we show the results of the numerical analysis based on a numerical code RTEQ (Ring Trap EQuilibrium) developed by Furukawa\,\cite{Furukawa}.

\subsection{Setting and calculation model}
\label{subsec:result1}

\begin{figure}
    \begin{center}
    \includegraphics[scale=0.45]{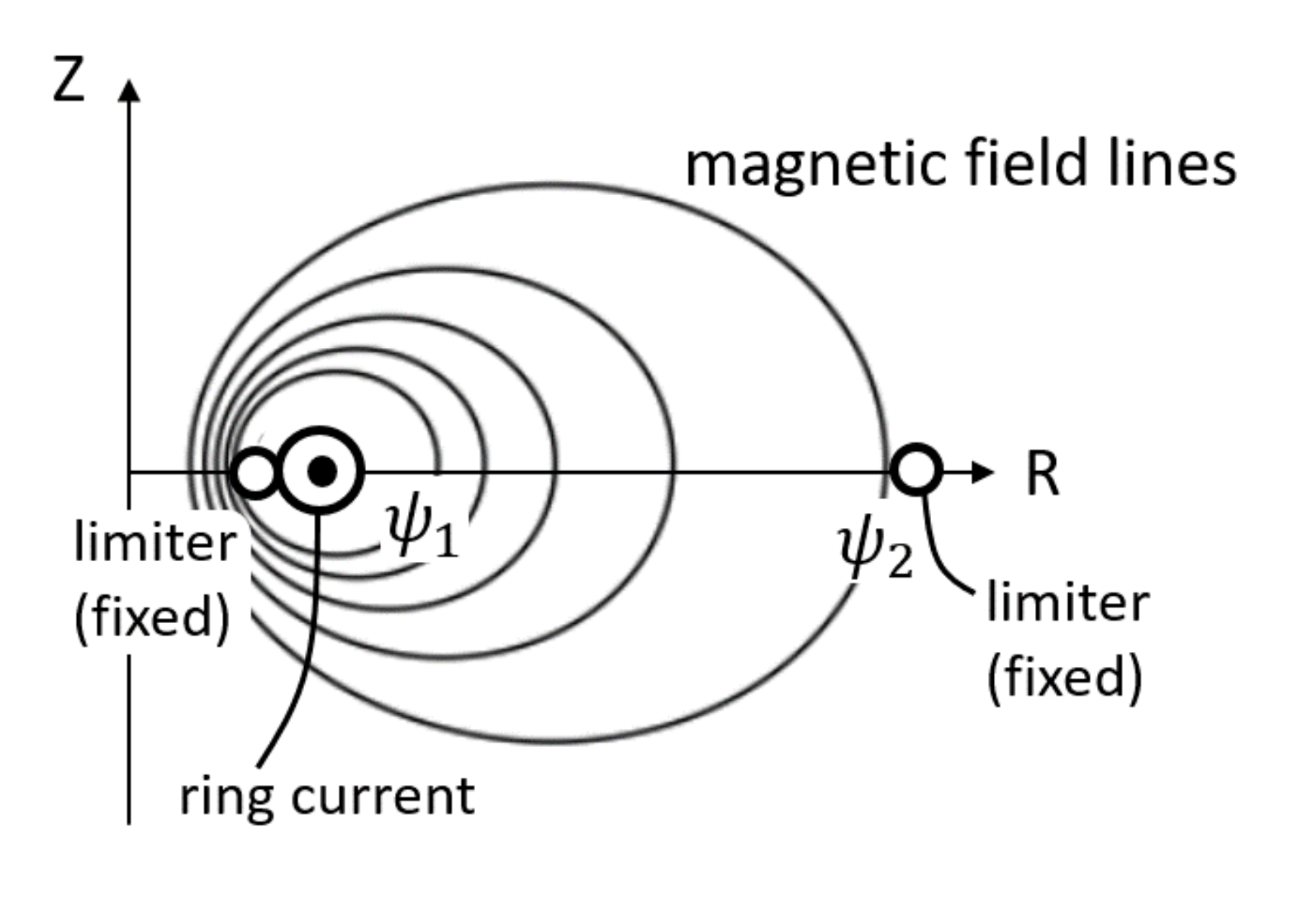}
    \end{center}
    \caption{
    \label{fig:graph1}
    Schematic of the 2D-plane considered in the calculation
    }
\end{figure}

As we assume axisymmetry, we consider the 2D-plane shown in fig.\,\ref{fig:graph1}. 
The magnetic field lines correspond to the $\psi$ contour, and $l=0$ on the magnetic field lines corresponds to the point where $z=0$ at the outer side of the ring current.
Here, $\psi_1$ and $\psi_2$ denote the magnetic field line that comes in contact with the fixed limiters that provide the boundary of the plasma.
The internal ring current (which is implemented by the levitated superconducting coil) generates the vacuum magnetic field.

Normalized with the typical length $L_1$, magnetic field $B_1$, pressure $p_1$, and magnetic flux $\Psi_1:=B_ 1 L _1^2$, the Grad--Shafranov equation is expressed as follows:
\begin{equation}
\check{\Delta}^*\check{\psi} = -\frac{\beta_o}{2}\check{r}^2\left.\frac{\partial\check{p}_\parallel}{\partial\check{\psi}}\right|_{\check{B}} - \frac{1}{\check{\sigma}}\check{\nabla}\check{\psi}\cdot\check{\nabla}\check{\sigma},\label{GS-3}
\end{equation}
where we apply (\ref{p_parallel}) and (\ref{p_perp}) as the parallel and perpendicular components of the pressure tensor, respectively, which are rewritten as
\begin{eqnarray}
\check{p}_\parallel(\psi,\,B)&&=\bar{p}(\check{\psi})\frac{\lambda_0}{\lambda_0+\frac{\check{B}_0(\psi)}{\check{B}}(1-\lambda_0)},\label{p_parallel3}\\
\check{p}_\perp(\psi,\,B)&&=\bar{p}(\check{\psi})\biggl(\frac{\lambda_0}{\lambda_0+\frac{\check{B}_0(\psi)}{\check{B}}(1-\lambda_0)}\biggr)^2.\label{p_perp3}
\end{eqnarray}
The boundary condition is $\psi=0$ at infinity. 
The equation contains two parameters: $\displaystyle\beta_o:=\frac{2\mu_0p_1}{B_1^2}$ and $\displaystyle\lambda_0:=\frac{T_{\perp 0}}{T_{\parallel 0}}$. 
Here, $\lambda_0$ represents the temperature anisotropy on $l=0$. 
Moreover, $\bar{p}(\check{\psi})$ corresponds to the normalized pressure in the isotropic case, which allows arbitrary functions of $\check{\psi}$, and $\check{B}_0(\psi):=\check{B}(0,\psi)$ is the magnetic field strength at $\ell=0$ as we defined in Sec.\,\ref{Maxwellian}.
Here, we assume
\begin{equation}
\bar{p}(\check{\psi})\propto-(\check{\psi}-\check{\psi_1})^P(\check{\psi}-\check{\psi}_2)^Q.
\end{equation}
Although we can treat $P$ and $Q$ as parameters, we fix $P=1,\, Q=1$ in this study for simplicity. 
Then, we obtain
\begin{equation}
\bar{p}(\check{\psi}):=-\frac{4(\check{\psi}-\check{\psi}_1)(\check{\psi}-\check{\psi}_2)}{(\check{\psi}_2-\check{\psi}_1)^2}.
\end{equation}

\subsection{Effects of anisotropic temperature on the equilibrium states}
\label{subsec:result2}
\begin{figure*}
    \begin{center}
    \includegraphics[scale=0.52]{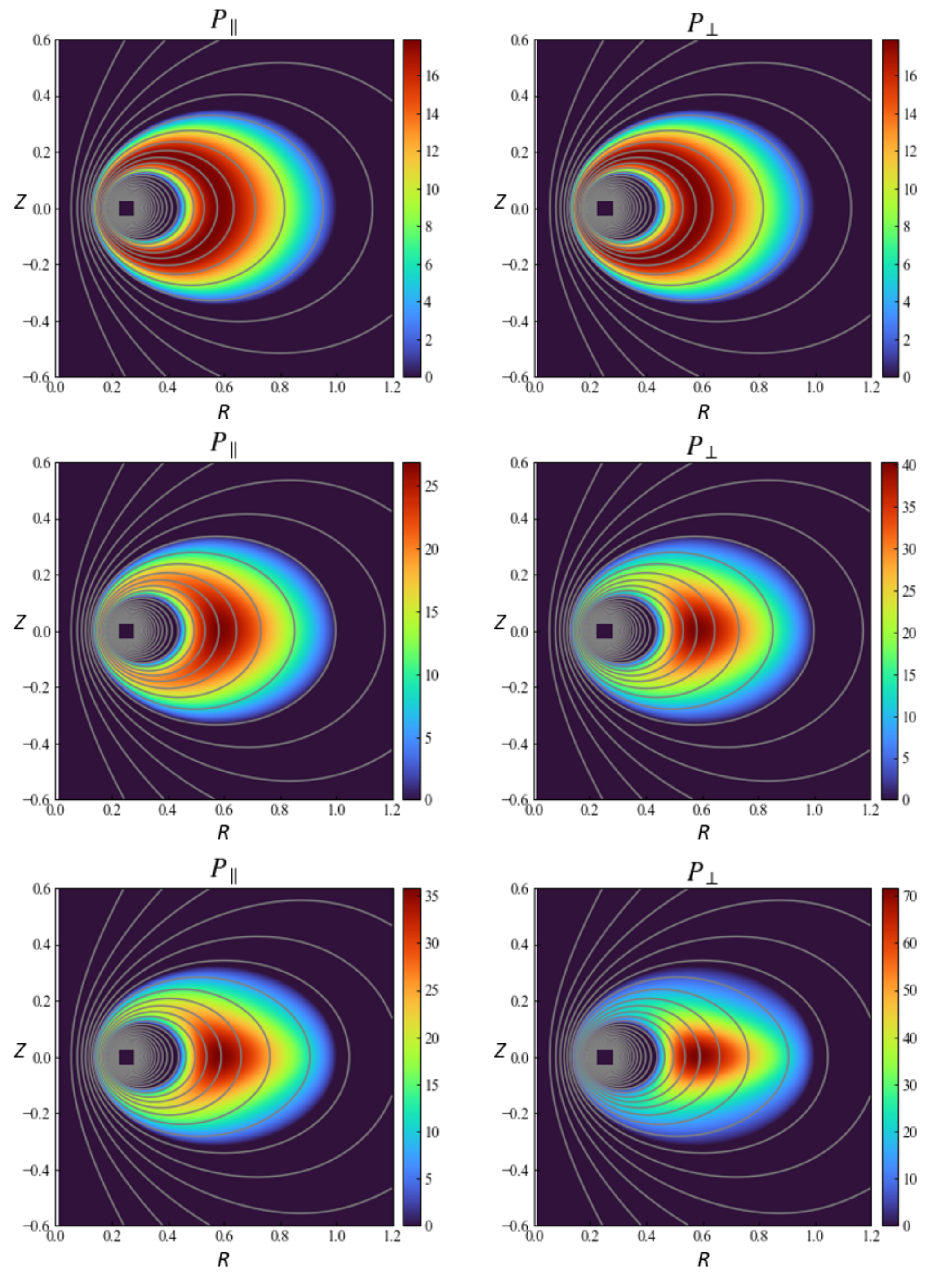}
    \end{center}
    \caption{
    \label{fig:graph2}
    Distributions of $p_\parallel$ and $p_\perp$, which correspond to the equilibrium states calculated for $\beta_o=4.5\times10^{-5}$ and (top) $\lambda_0=1.0$, (middle) $\lambda_0=1.5$, (bottom) $\lambda_0=2.0$
    }
\end{figure*}

\begin{figure*}
    \begin{center}
    \includegraphics[scale=0.5]{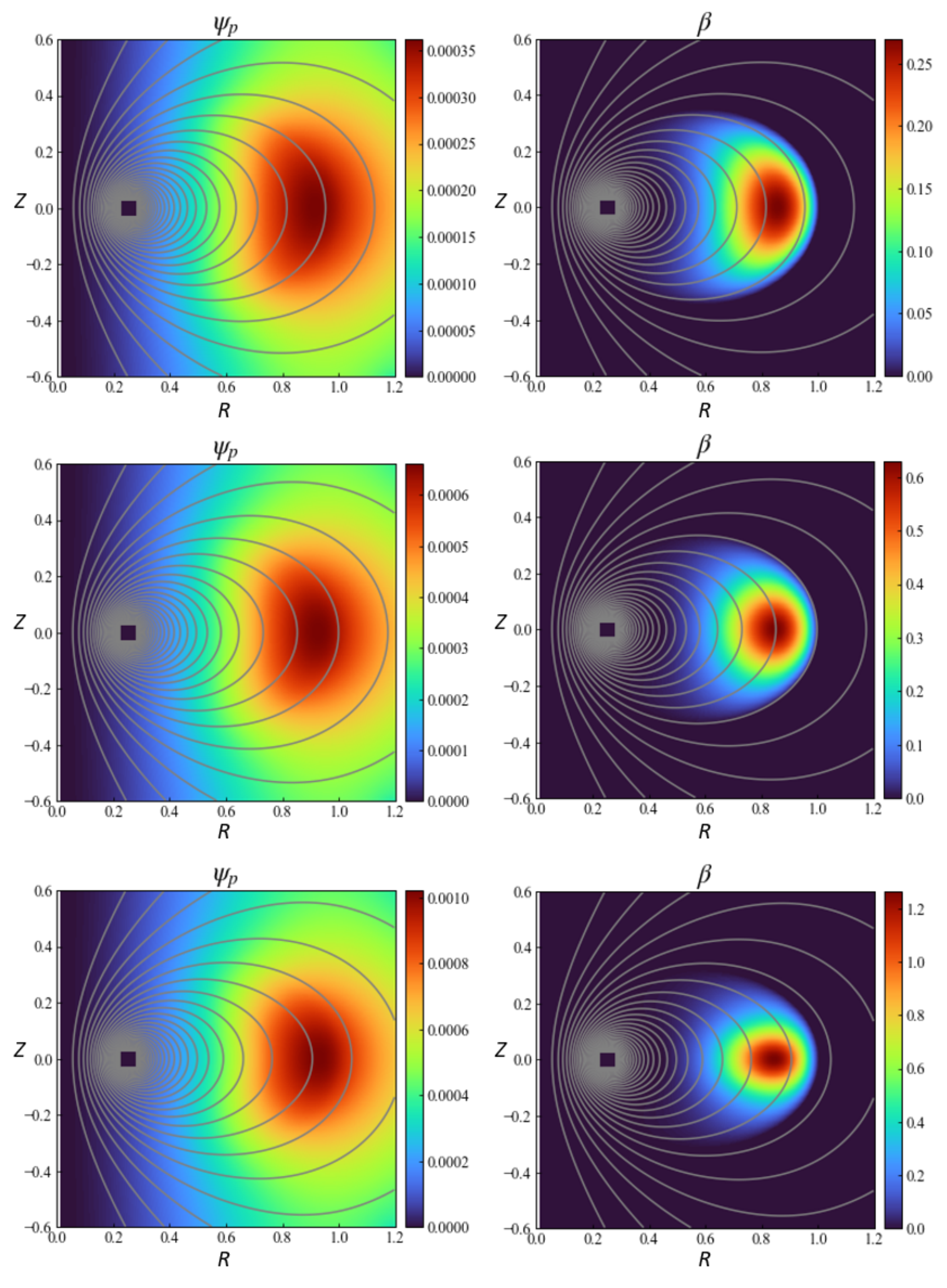}
    \end{center}
    \caption{
    \label{fig:graph3}
    Distributions of $\psi_p$ and $\beta$, which correspond to the equilibrium states calculated for $\beta_o=4.5\times10^{-5}$ and (top) $\lambda_0=1.0$, (middle) $\lambda_0=1.5$, (bottom) $\lambda_0=2.0$
    }
\end{figure*}

First, we analyze the cases where $\lambda_0\geq 1$. 
We show the distributions of $p_\parallel$ and $p_\perp$ in fig.\,\ref{fig:graph2} and those of $\psi_p$ and $\beta$ in fig.\,\ref{fig:graph3}, which correspond to the equilibrium states calculated for $\beta_o=4.5\times10^{-5}$ and (top) $\lambda_0=1.0$, (middle) $\lambda_0=1.5$, (bottom) $\lambda_0=2.0$.
Here, $\displaystyle\beta:=\frac{2\mu_0p}{B^2}$ and $\psi_p:=\psi-\psi_v$, where $\psi_v$ denotes the flux function associated with a vacuum magnetic field.
As shown in fig.\,\ref{fig:graph2} when $\lambda_0=1$, the distributions of pressure become isotropic ($\check{p}_\parallel=\check{p}_\perp=\bar{p(\psi)}$). 
Therefore, $p_\parallel$ and $p_\perp$ are constant along the magnetic field lines. 
On the other hand, when $\lambda_0>1$, $p_\parallel$ and $p_\perp$ are higher at the outer side of the ring current near $z=0$. 
Moreover, as $\lambda_0$ increases, the distributions are more concentrated at the outer side of the ring current near $z=0$, and the maximum local values of $p_\parallel$ and $p_\perp$ become higher. 
Even for $\psi_p$ and $\beta$, the distributions are more concentrated at the outer side of the ring current near $z=0$, and the maximum local values become higher as $\lambda_0$ increases; see fig.\,\ref{fig:graph3}.

\begin{figure*}
    \begin{center}
    \includegraphics[scale=0.52]{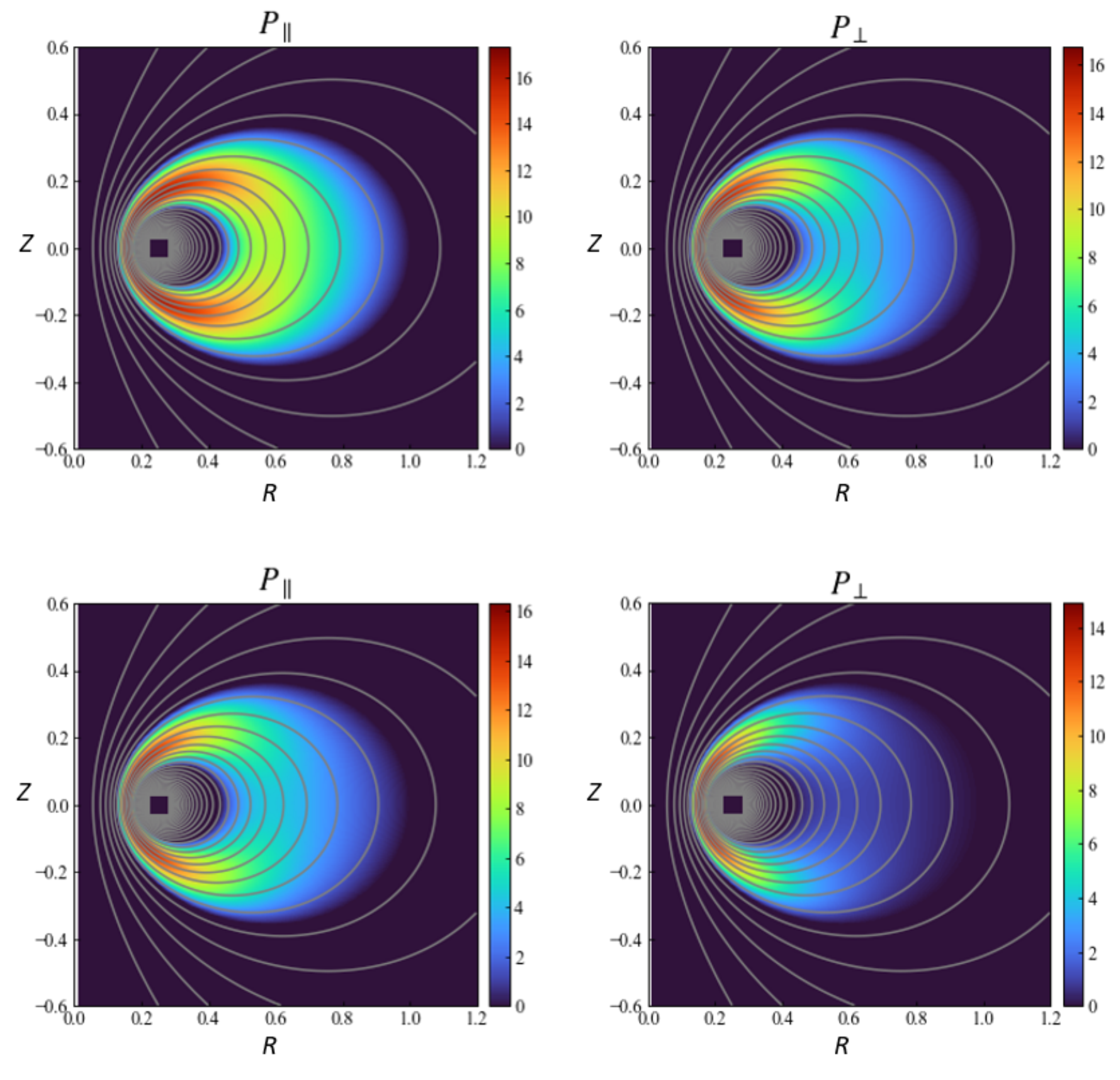}
    \end{center}
    \caption{
    \label{fig:graph4}
    Distributions of $p_\parallel$ and $p_\perp$, which correspond to the equilibrium states calculated for $\beta_o=4.5\times10^{-5}$ and (top) $\lambda_0=0.50$, (bottom) $\lambda_0=0.25$ 
    }
\end{figure*}

\begin{figure*}
    \begin{center}
    \includegraphics[scale=0.48]{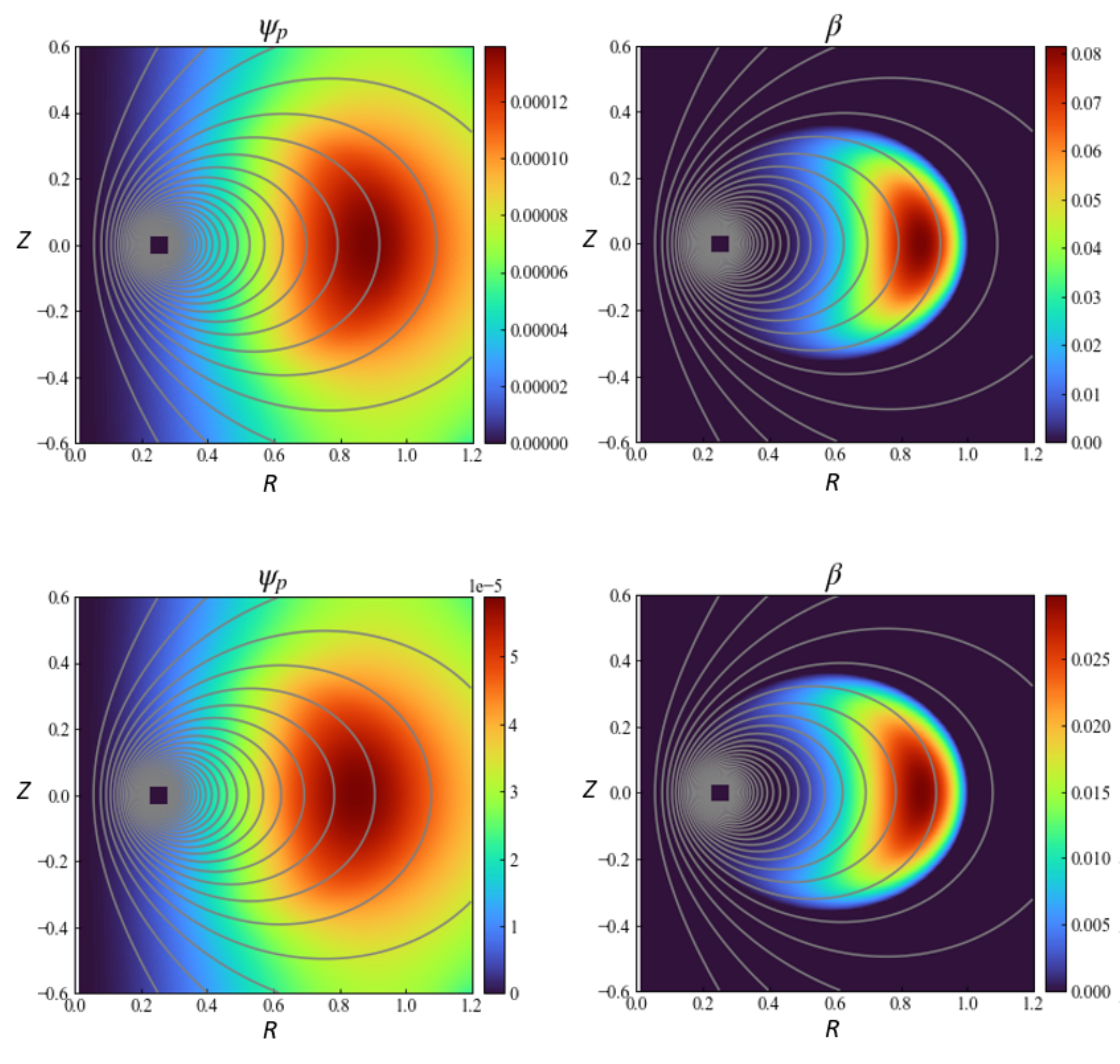}
    \end{center}
    \caption{
    \label{fig:graph5}
    Distributions of $\psi_p$ and $\beta$, which correspond to the equilibrium states calculated for $\beta_o=4.5\times10^{-5}$ and (top) $\lambda_0=0.50$, (bottom) $\lambda_0=0.25$
    }
\end{figure*}

Next, we analyze the cases where $\lambda_0<1$.
We show the distributions of $p_\parallel$ and $p_\perp$ in fig.\,\ref{fig:graph4} and those of $\psi_p$ and $\beta$ in fig.\,\ref{fig:graph5}, which correspond to the equilibrium states calculated for $\beta_o=4.5\times10^{-5}$ and (top) $\lambda_0=0.50$, (bottom) $\lambda_0=0.25$. 
In line with the cases in which $\lambda\geq 1$, $p_\parallel$ and $p_\perp$ are higher at the inner side of the ring current near $z=0$. 
Moreover, as $\lambda_0$ becomes smaller, the distributions are more concentrated at the inner side of the ring current near $z=0$, but the maximum local values of $p_\parallel$ and $p_\perp$ become lower (see fig.\,\ref{fig:graph4}). 
Regarding $\psi_p$ and $\beta$, the distributions are concentrated at the outer side of the ring current, and the maximum local values become lower as $\lambda_0$ becomes smaller; see fig.\,\ref{fig:graph5}.

\begin{figure}
    \begin{center}
    \includegraphics[scale=0.42]{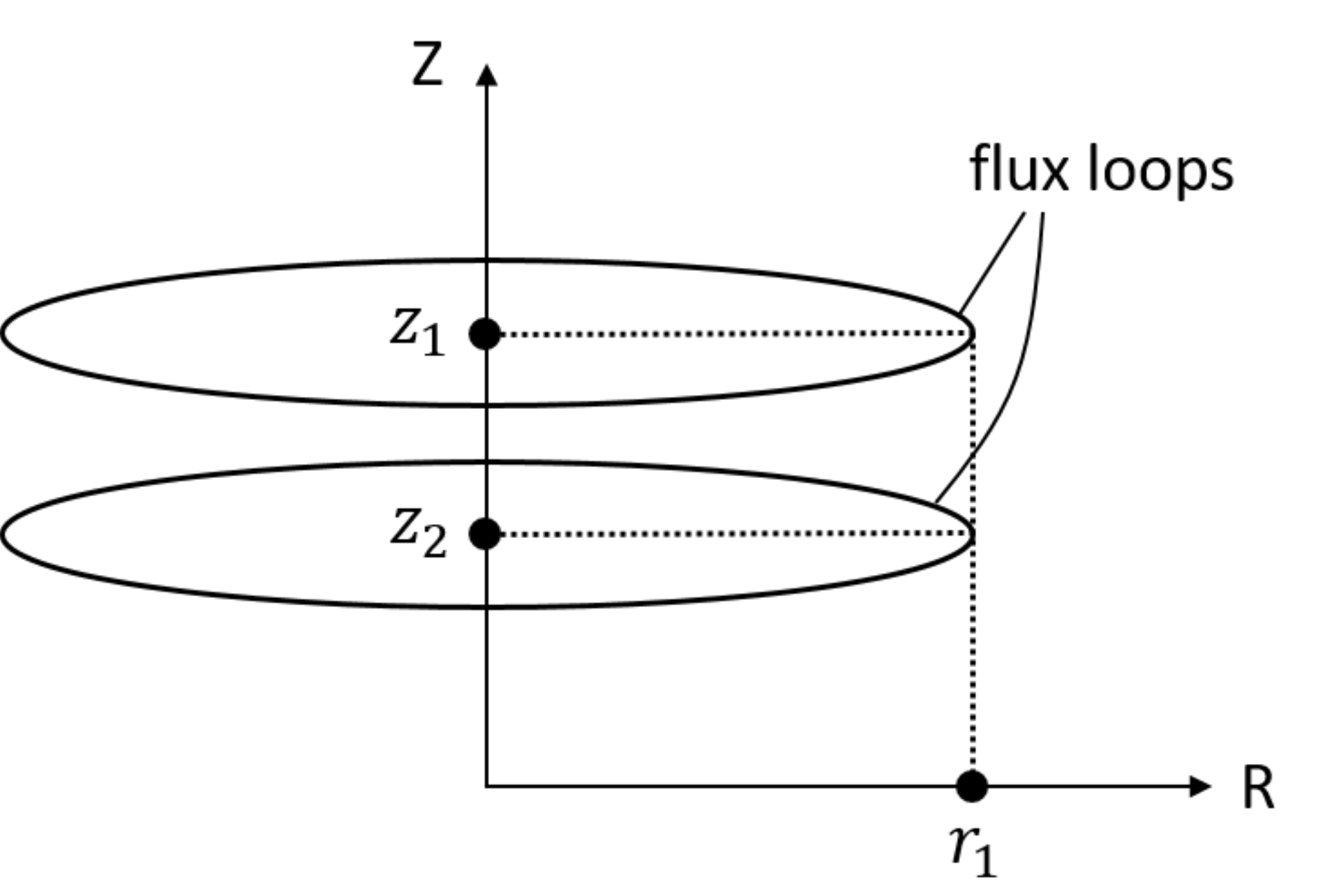}
    \end{center}
    \caption{
    \label{fig:graph6}
    Schematic of the flux loops
    }
\end{figure}

\begin{figure}
    \begin{center}
    \includegraphics[scale=0.8]{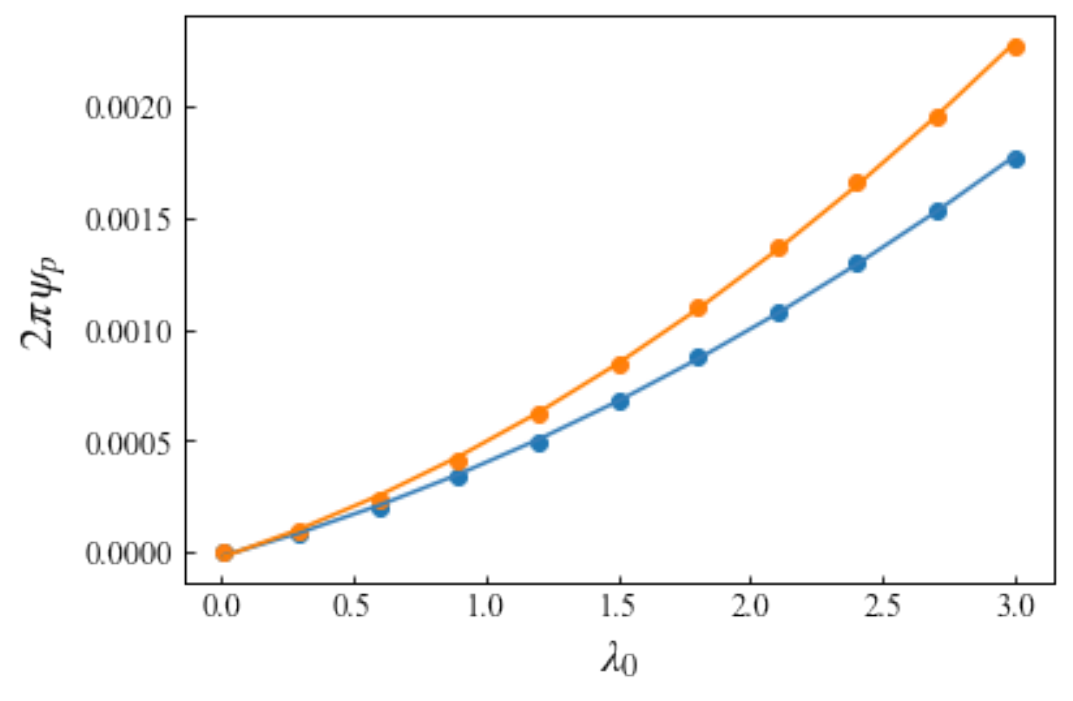}
    \end{center}
    \caption{
    \label{fig:graph7}
    Relation between $\lambda_0$ and $\psi_p$ where $\beta_o=1.2\times10^{-5}$
    }
\end{figure}

\begin{figure}
    \begin{center}
    \includegraphics[scale=0.8]{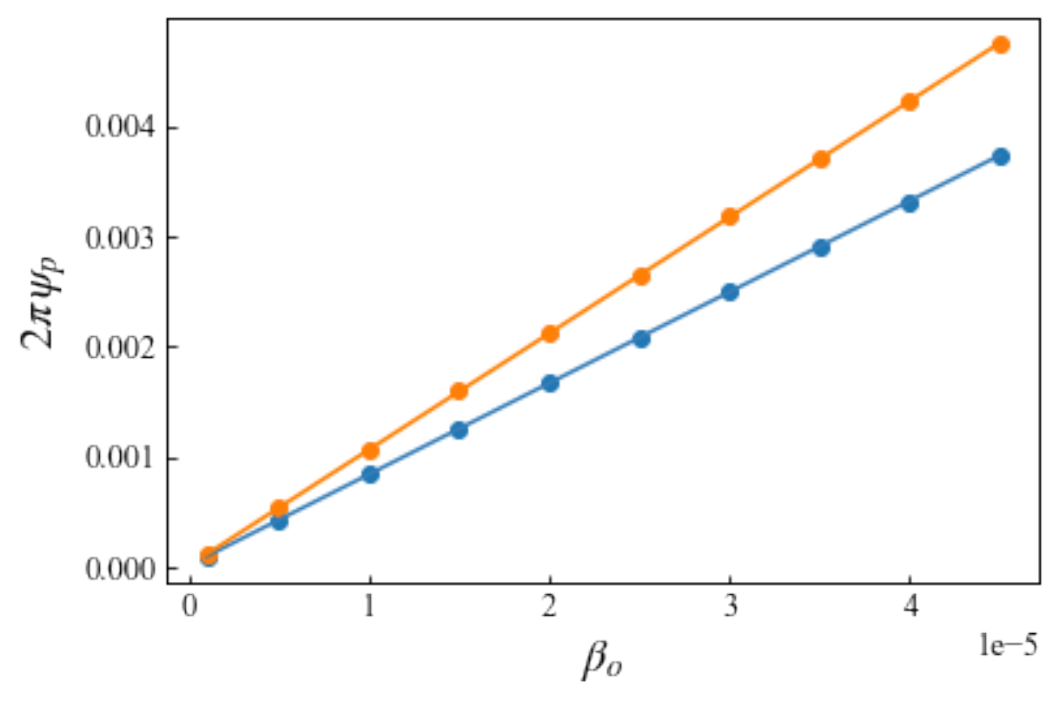}
    \end{center}
    \caption{
    \label{fig:graph8}
    Relation between $\beta_o$ and $\psi_p$ where $\lambda_0=2.0$ 
    }
\end{figure}

Finally, we show the basic idea of applying our theoretical model to experimental studies, which will be future works. 
When we apply the model to experiments, we have to determine the value of $\beta_o$ and $\lambda_0$. 
By measuring the magnetic induction on a toroidal flux loop (fig.\,\ref{fig:graph6}), we can evaluate $\psi_p$ at the corresponding coordinate $r$ and $z$. 
By fitting the numerical solutions of the model and experimental $\psi_p$ at multiple points, we can estimate the parameters (see fig.\,\ref{fig:graph7}-\ref{fig:graph8}).
Here, we set $r_1=1.01,\,z_1=0.35,\,z_2=0.20$.

\clearpage
\section{Conclusion}
\label{sec:conclusion2}
To describe the high-beta equilibrium of a magnetospheric plasma, we need a consistent relation between the magnetic field (to be modified by the current in the plasma) and the phase-space distribution function (to be influenced by the magnetic field).
The former is dictated by the field equation, and the latter is dictated by kinetic theory.
We can use the (generalized) Grad--Shafranov equation as the field equation that determines the magnetic flux function for a given magnetization (diamagnetic) current.
However, the Grad--Shafranov equation has an additional (in fact, essential) implication, that is, the internal relation between the magnetic field and the magnetization current, which is imposed by the magneto-fluid force--balance relation.
Therefore, we need to find a special class of distribution functions that does not create inconsistencies with the macroscopic magneto-fluid model.
In the present study, we showed that the ``thermal equilibrium'' on the topologically constrained phase space (foliated by the adiabatic invariant $\mu$) is suitable for the generalized Grad--Shafranov equation, which is not only amenable, but also definitive for the functional form of the pressure tensor.
However, we have left the flux function $\psi$ as a free parameter that can control the ``radial'' profile of the pressure tensor.
Hasegawa\,\cite{hasegawa1987} suggested that $\partial_\psi f^*=0$, because $P_\theta\sim \psi$ is the most fragile constant influenced by low-frequency ($\sim$ drift frequency) perturbation.
In fact, we observe the ``inward diffusion'' of particles, consistent with the relaxation toward $\partial_\psi f^*=0$\,\cite{garnier2009,kawazura2015}.
However, in a real system, the boundaries (located both inside and outside the confinement domain) deform the distribution from the ideal one.
In the present study, we maintained $\psi$ as an experimental parameter to model the nonequilibrium property.
The theoretical model constructed in the present work will be useful to analyze the experimental data of dipole systems or satellite data of planetary magnetospheres. 
The function $A(\psi)$ will characterize the non-equilibrium property of the real system in comparison with the simple model of the relaxed state\,\cite{hasegawa1987}. 
The two parameters $\beta_o$ (measuring the beta) and $\lambda_0$ (characterizing the temperature anisotropy) can be determined, in experiments, by measuring the variation of magnetic flux $\psi_p$ using flux loops. 
We observe a variety of density and pressure profiles depending on plasma parameters\,\cite{saitoh2014}. 
Detailed experimental analysis will be discussed in future work.

\section*{Acknowledgment}
The authors thank Professor M. Furukawa and Mr. K. Ueda for their support in the numerical calculations.
This study was supported by JSPS KAKENHI (grant number 17H01177).

\appendix
\section{Grand canonical ensemble}\label{appendix:GCE}
The function form of (\ref{f-star1}) implicitly assumes maximum entropy states under the appropriate constraints formulated in \cite{yoshida2014self}. 

When we consider the grand canonical ensemble determined by the Casimir (magnetic moment) $C=\int\mu f^*\rmd^nz$ in addition to the total particle number $N=\int f^*\rmd^nz$ and total energy $E=\int Hf^*\rmd^nz$, the equilibrium state in which entropy $S=-\int f^*\log f^*\rmd^nz$ is maximized is calculated as
\begin{equation}
    \delta(S-\alpha N-\beta E-\gamma C)=0,\label{ELeq}
\end{equation}
which yields a Boltzmann distribution
\begin{equation}
    f^*=Z^{-1}\exp(-\beta H-\gamma\mu), 
\end{equation}
where $Z:=\exp(\alpha+1)$ is the normalized factor and $\alpha,\,\beta,\,\gamma$ are Lagrange multipliers.
The Euler-Lagrange equation of (\ref{ELeq}) is formally the same as that of a different variational principle $\delta(E+\alpha_1 S+\alpha_2 N+\alpha_3 C)=0$, which may be interpreted as the “Energy-Casimir” functional used in the stability theory of non-canonical Hamiltonian systems (entropy $S$ may be regarded as a Casimir of the Vlasov Lie-Poisson algebra). 
Here, we invoke (\ref{ELeq}) to define the statistical equilibrium on the “energy shell” that is constrained by $C$.

In our formulation, we also consider $\psi$ as a constant of motion. 
Then, $Z,\,\beta$ and $\gamma$ can be the functions of $\psi$. 
However, in this study, we consider $\beta$ to be a constant in (\ref{f-star1}) for simplicity. 

When we do not consider the magnetic moment as a constraint, the equilibrium state obtained via entropy maximization changes drastically. 
The distribution function is obtained as
\begin{equation}
    f^*=Z^{-1}\exp(-\beta H),
\end{equation}
which is equivalent to the distribution function obtained by $T_{\parallel 0}=T_{\perp 0}=T_0$ in (\ref{f-star1}). 
The distribution function yields the pressure as
\begin{equation}
    p_\parallel=p_\perp=A(\psi)\biggl(\frac{2\pi}{m}\biggr)^{\frac{3}{2}}T_{\parallel 0}^{\,\frac{5}{2}},
\end{equation}
which is isotropic and the function of $\psi$ (not $\psi$ and $B$). 
This simple exercise shows that the factor that causes the nontrivial structure along the magnetic field at the same time as entropy maximization is the constraint of the magnetic moment in the view of statistical mechanics.
\bibliographystyle{unsrt}
\bibliography{equilibrium}
\end{document}